# Investigation of Machine Learning-based Coarse-Grained Mapping Schemes for Organic Molecules

Dimitris Nasikas

*Institute of Nanoscience and Nanotechnology, National Center for Scientific Research "Demokritos", Athens, Greece & School of Chemical Engineering, National Technical University of Athens, Athens, Greece*, nasidimi@gmail.com

Eleonora Ricci[*]

*Institute of Informatics and Telecommunications & Institute of Nanoscience and Nanotechnology, National Centre for Scientific Research "Demokritos", Athens, Greece*, e.ricci@inn.demokritos.gr

George Giannakopoulos

*Institute of Informatics and Telecommunications, National Centre for Scientific Research "Demokritos", Athens, Greece & Scify P.N.P.C., Athens, Greece*, ggianna@iit.demokritos.gr

Vangelis Karkaletsis

*Institute of Informatics and Telecommunications, National Centre for Scientific Research "Demokritos", Athens, Greece*, vangelis@iit.demokritos.gr

Doros N. Theodorou

*School of Chemical Engineering, National Technical University of Athens, Athens, Greece*, doros@chemeng.ntua.gr

Niki Vergadou[*]

*Institute of Nanoscience and Nanotechnology, National Centre for Scientific Research "Demokritos", Athens, Greece*, n.vergadou@inn.demokritos.gr

**ABSTRACT**

Due to the wide range of timescales that are present in macromolecular systems, hierarchical multiscale strategies are necessary for their computational study. Coarse-graining (CG) allows to establish a link between different system resolutions and provides the backbone for the development of robust multiscale simulations and analyses. The CG mapping process is typically system- and application-specific, and it relies on chemical intuition. In this work, we explored the application of a Machine Learning strategy, based on Variational Autoencoders, for the development of suitable mapping schemes from the atomistic to the coarse-grained space of molecules with increasing chemical complexity. An extensive evaluation of the effect of the model hyperparameters on the training process and on the final output was performed, and an existing method was extended with the definition of different loss functions and the implementation of a selection criterion that ensures physical consistency of the output. The relationship between the input feature choice and the reconstruction accuracy was analyzed, supporting the need to introduce rotational invariance into the system. Strengths and limitations of the approach, both in the mapping and in the backmapping steps, are highlighted and critically discussed.

---

[*] To whom correspondence should be addressed

# 1 Introduction

Coarse-Grained (CG) molecular simulations [1] allow to access larger length and timescales, thereby expanding the reach of molecular modelling, according to the coarse graining level adopted. Macromolecules, such as polymers, are indeed systems that stand to greatly benefit from the utilization of a CG molecular description [2], to overcome the time and length scale limitations of a full atomistic representation, at the cost of losing part of the fine-grain chemical detail [3,4].

Choosing a CG level entails averaging out a certain number of high-resolution degrees of freedom (*e.g.* atoms) into a single interaction site, or coarse-grained bead, *i.e.* choosing a mapping scheme. Subsequently, it is necessary to represent the potential energy landscape of the high-resolution system in the chosen lower dimensional CG space, *i.e.* develop a CG force field. Oftentimes, a third step comes into play, when it is necessary to backmap the CG system to the atomistic level of detail. In this work, we focus on the choice of CG mapping and on the back mapping steps.

A CG mapping is typically chosen using chemical or physical intuition, symmetry considerations, and convenience. The CG mapping is generally related to the specific problem and the scale of the processes under study and therefore there is no unique way to coarse-grain. The application of Machine Learning techniques to this task could help in optimizing the CG mapping process and in contributing to the development of systematic reverse mapping routes.

In this work, we explored the application of Variational Autoencoders (VAE) [5] to generate CG mapping, by interpreting the latent space variables of the network as CG coordinates. The size of the latent dimension is used to specify the CG level, and the target property in training the VAE is the reconstruction of the atomistic coordinates. This approach was demonstrated on a bulk system of polyethylene (PE) oligomers ($C_{24}H_{50}$) [6] and also in the case of a bulk ionic liquid [7]. In the present work [8], we extended the existing method by adding various expressions for the loss definition and applying physically consistent selection criteria to the extracted CG mappings. Ethane, ethylbenzene, and the monomeric units of cellulose triacetate (CTA) and polymer of intrinsic microporosity PIM-1 were chosen as test cases, to investigate the method on systems of increasing complexity.

# 2 METHODOLOGY

## 2.1 Training data generation

The data used for training the model were generated by running molecular simulations at the atomistic level on the various systems, the chemical structures of which are shown in Fig. 1. For each system, 104 isolated molecule configurations were generated, sampled from microcanonical molecular dynamics (NVE MD) trajectories of 1 ns, with integration step 1 fs. For the simulation of ethane and ethylbenzene, the Dreiding force field was chosen [9], while the Glycam06 force field was chosen for CTA [10] and the polymer consistent force field [11] for PIM-1.

## 2.2 Variational Autoencoders as CG tools

The Cartesian coordinates of the data generated from the aforementioned molecular simulations were used as input to train Variational Autoencoders (VAE). An Autoencoder is a type of neural network that is typically trained to reproduce its input data. Its middle layers have lower dimension compared to the input/output ones.

Therefore, owing to the presence of this bottleneck in the network architecture, the VAE learns to reconstruct the data from a compressed latent representation. The model consists of two parts, an Encoder, that in the present case maps the atomistic input to the CG latent representation, and a Decoder, that maps from the latent representation back to the atomistic level. In this work, one intermediate layer was used.

The choice of a VAE as a coarse-graining tool is based on its ability to reduce the dimensionality of the input data, by projecting them to a lower dimensional space, and then reconstruct the initial data, by projecting them back to the initial space. Thus, in a way it can model coarse-graining and reverse mapping at the same time.

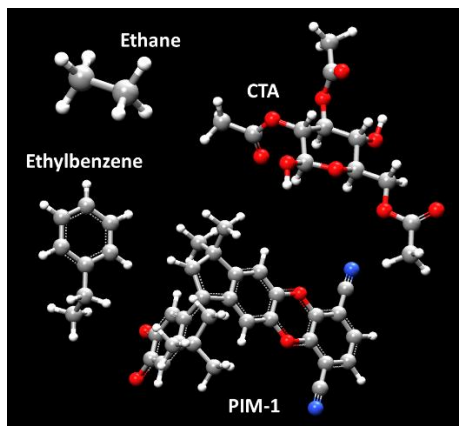

**Figure 1:** Molecules analyzed in this study.
[Elements list: C :: grey, H :: white , O :: red , N :: blue.]

The projection to the lower dimensional space through the Encoder should be done in such a way that sufficient information is preserved to reconstruct the initial atomistic coordinates from the latent space. Through the Encoder, we are searching for a tensor that can project the atomistic coordinates to a desired coarse-grained space and through the Decoder we are searching for a tensor that can project the latent vector back to the initial atomistic coordinates. Let $E$ be the Encoder's tensor, which performs a projection as indicated in the following:

$$E: \mathbb{R}^{3n} \to \mathbb{R}^{3N}, n, N \in \mathbb{N}^*, n > N.$$

Where $n$ is the number of atoms $j$ in the full atomistic representation and $N$ is the number of coarse-grained moieties $i$ that substitute the atoms. $E$ can be written in a matrix form:

$$E = \begin{pmatrix} E_{11} & E_{12} & \cdots & E_{1n} \\ E_{21} & E_{22} & \cdots & E_{2n} \\ . & . & \cdots & . \\ E_{N1} & E_{N2} & \cdots & E_{Nn} \end{pmatrix}$$

Let $x$ be the initial atomistic coordinates and $z$ the coordinates in the coarse-grained space. The *i*-th coarse-grained moiety's coordinates can be described by the following relationship:

$$z_i = \sum_{j=1}^{n} E_{ij} \cdot x_j \, , \in \mathbb{R}^3$$

Some important characteristics of the $E$ matrix are that $E_{ij} \geq 0$ and that its rows are normalized to 1. It is necessary to perform this step in order to interpret $E_{ij}$ as the coefficients indicating the participation of each atom to each coarse grained moiety [12]:

$$\sum_{j=1}^{n} E_{ij} = 1$$

The $E$ matrix is obtained here from the normalization by row of the weight matrix of the Encoder. Therefore, the atomistic coordinates that are used as input data, are connected to the Encoder's neurons via weights that are stored in $E$. The weight values $E_{ij}$ are expressed as a function of stochastic parameters $C_{ij}$ as follows:

$$E_{ij} = \frac{C_{ij}}{\sum_{j=1}^{n} C_{ij}}$$

The $C_{ij}$ parameters have a probabilistic interpretation. Specifically, $C_{ij}$ expresses the probability of assignment of an atom $j$ to a coarse-grained moiety $i$. Thus, the $C_{ij}$ coefficients are normalized when summed over $j$. An assignment matrix $C$ can be defined. Each column of the $C$ matrix contains the probabilities of each atom $j$ to belong to a coarse-grained moiety $i$. Each atom is assigned to the moiety with the highest assignment probability

$$\text{moiety of atom } j = \text{index}\left[\max\left(C_{1j}, C_{2j}, \dots, C_{Nj}\right)\right]$$

Moreover, $C_{ij}$ are sampled from the Gumbel-Softmax distribution where a reparametrization is applied to ensure that the network is differentiable and allow backpropagation to take place during the training process [13]. The above is known as the Gumbel-Softmax reparametrization trick.

$$c_{ij} = \frac{\exp\left[\frac{\log \phi_{ij} + g_{ij}}{\tau}\right]}{\sum_{i=1}^{N} \exp\left[\frac{\log \phi_{ij} + g_{ij}}{\tau}\right]}$$

The $\phi_{ij}$ parameters correspond here to the encoder's weight, and so to the probability of an atom $j$ to belong to a coarse-grained moiety $i$, and $g_{ij}$ is a parameter sampled from the Gumbel distribution. Finally, $\tau$ is a "temperature" parameter whose value follows a stepwise decrease until the end of the training making the distributions of atoms to coarse-grained moieties narrower, approaching the limit of a one-hot encoding.

Following the projection to the coarse-grained level, the Decoder uses the latent vector produced by the Encoder to reconstruct the full atomistic space. Let $\boldsymbol{D}$ be the Decoder's tensor. Mathematically it can be described as:

$$\boldsymbol{D}: \mathbb{R}^{3N} \to \mathbb{R}^{3n}, n, N \in \mathbb{N}^*$$

And $\boldsymbol{D}$ can be expressed in a matrix form:

$$\boldsymbol{D} = \begin{pmatrix} D_{11} & D_{12} & \dots\dots & D_{1N} \\ D_{21} & D_{22} & \dots\dots & D_{2N} \\ . & . & \dots\dots & . \\ D_{n1} & D_{n2} & \dots\dots & D_{nN} \end{pmatrix}$$

The reconstructed vectors are given by the following relation:

$$\boldsymbol{x}_j' = \sum_{i=1}^{N} D_{ij} \cdot \boldsymbol{z}_i, \; \boldsymbol{x}_j' \in \mathbb{R}^3$$

In summary, after a successful VAE's training, we obtain a coarse-grained mapping and a reverse mapping at the same time.

## 2.3 Loss Function

In order to train the VAE, a loss function should be defined. For VAEs the most typical choice is an error metric, such as MSE, calculated from the difference between the input x and the reconstructed output x':

$$\mathcal{L}_{\text{rec}} = \frac{1}{N} \mathbb{E}_{\mathbf{x}}[(\mathbf{x}' - \mathbf{x})^2] \quad (1)$$

In addition to the reconstruction, Wang et al. [6] proposed the addition of an instantaneous force regularization term, to promote the learning of mapping schemes that minimize the average force in the CG space, thus yielding a smoother potential energy hypersurface, to facilitate the subsequent learning of a CG force field [14]. The loss function is thus defined as:

$$\mathcal{L}_{\text{total}} = \mathcal{L}_{\text{rec}} + \rho \mathcal{L}_{\text{forces}} = \frac{1}{N} \mathbb{E}_{\mathbf{x}}[(\mathbf{x}' - \mathbf{x})^2 + \rho \cdot F^2] \quad (2)$$

where ρ acts as a weighting factor that controls the balance between the two components in the loss function. Indeed, depending on the choice of units, the instantaneous average force can be orders of magnitude larger than the reconstruction part of the loss, leading to extreme behaviors from the model during the optimization process, which will be discussed in Section 3. Choosing an appropriate value for the hyperparameter ρ can mitigate this effect. However, the choice of an optimal value for ρ is heavily dependent on the chemical system examined, which acts as an obstacle for a transferable application of the method. Therefore, an alternative loss function is proposed here, based on the observation that the optimal value of ρ to avoid degenerate behavior was observed to be the one that brings both terms to similar orders of magnitude. To achieve this in a system-independent manner, the reconstruction as well as the forces part of the loss, are normalized. The normalization is performed by dividing each term by the maximum value it assumed until the current epoch of the optimization, and it can be written as:

$$\mathcal{L}_{\text{norm},K} = (1-a) \cdot \frac{\mathcal{L}_{\text{rec}}}{\max_{k=1,K} \mathcal{L}_{\text{rec},k}} + a \frac{\mathcal{L}_{\text{forces}}}{\max_{k=1,K} \mathcal{L}_{\text{forces},k}} \quad (3)$$

Where $\alpha \in [0,1]$, is an hyperparameter that regulates the balance between the two terms of the loss function and K is the current optimization epoch. In each denominator, the maximum $\mathcal{L}_{\text{rec}}$ and $\mathcal{L}_{\text{forces}}$ value recorded until the current epoch of the optimization process is chosen.

For the loss function expressed with Eq. (2) and Eq. (3), two different schemes were examined. In the first one, the forces' part is activated only after the reconstruction has reached a predefined threshold value, akin to what is suggested by Wang et al. [6]. The purpose of this criterion is to proceed with the forces part of the optimization only after the optimization of the reconstruction part has reached a desired level. In the second case, both terms were computed from the beginning. No systematic effect resulted from these two implementations, therefore it was chosen to keep both terms from the beginning for simplicity.

## 2.4 Mapping connectivity to the CG space

The determination of the connections (CG bonds) between the coarse-grained moieties provides critical information about a system's configuration in the coarse-grained space, especially for complex systems with multi-bead representations. To map the connectivity from the atomistic to the CG space, as a first step, an adjacency matrix (an undirected graph) is built to represent the connections between the atoms in the full atomistic configuration. Each atom is assigned to a unique identity, which is an integer number. Let $\boldsymbol{A_C}$ be the adjacency matrix. If an atom k is connected to an atom $l$ then $A_{kl} = A_{lk} = 1$, otherwise $A_{kl} = A_{lk} = 0$. For example, for a simple ethane molecule ($C_2H_6$) with atom IDs H:1, C:2, H:3, H:4, H:5 C:6, H:7 and H:8 since every hydrogen atom is connected to a carbon atom and the two carbon atoms are connected between them the adjacency matrix is:

$$\boldsymbol{A_C} = \begin{pmatrix} 0 & 1 & 0 & 0 & 0 & 0 & 0 & 0 \\ 1 & 0 & 1 & 1 & 0 & 1 & 0 & 0 \\ 0 & 1 & 0 & 0 & 0 & 0 & 0 & 0 \\ 0 & 1 & 0 & 0 & 0 & 0 & 0 & 0 \\ 0 & 0 & 0 & 0 & 0 & 1 & 0 & 0 \\ 0 & 1 & 0 & 0 & 1 & 1 & 1 & 1 \\ 0 & 0 & 0 & 0 & 0 & 1 & 0 & 0 \\ 0 & 0 & 0 & 0 & 0 & 1 & 0 & 0 \end{pmatrix}$$

After training the VAE model, from the assignment matrix $\boldsymbol{C}$, we deduce to which moiety each atom is assigned. If there is a connection between two atoms $k$ and $l$ that are assigned to two different moieties $s$ and $t$, i.e., if $A_{kl} = 1$, then a connection between the moieties $s$ and $t$ is also present. Let $\boldsymbol{CG_C}$ be the coarse-grained level connectivity matrix. When there is a connection between two moieties $s$ and $t$ then $CG_{C,st} = CG_{C,ts} = 1$, else $CG_{C,st} = CG_{C,ts} = 0$. For the ethane example, if we end up with two moieties with the first group containing the atoms with IDs 1,2,3 and 4 and the second the atoms with IDs 5,6,7 and 8, then, because of the connection between the carbons that belong in two different moieties, the $\boldsymbol{CG_C}$ matrix describing the connections between the moieties can be written as:

$$\boldsymbol{CG_C} = \begin{pmatrix} 0 & 1 \\ 1 & 0 \end{pmatrix}$$

## 3 RESULTS AND DISCUSSION

### 3.1 Hyperparameters optimization

A critical aspect for the training of an ML model is hyperparameter tuning. In the present VAE model, the values of several hyperparameters must be set, namely:
- learning rate
- decay ratio
- number of training epochs
- Gumbel-Softmax, reparameterization scaling factor, $\tau$
- batch size
- force regularization weight, $\rho$ (un-normalized loss) or $\alpha$ (normalized loss)
- number of nodes in the latent dimension, $N_{cg}$

In all tests, the initial and final values used for $\tau$ were 4 and $10^{-1}$, respectively. An early stopping criterion was implemented to consider training concluded if, for 10 consecutive epochs, the loss function did not vary by more than $10^{-2}$. Therefore, the number of epochs used for training was either determined by the early stopping criterion or by an upper threshold of $5 \times 10^3$. With these assumptions, five hyperparameters remain, and a grid search was conducted to determine their effect on the final value of the loss, on the mapping, and on the reconstruction. This preliminary investigation of hyperparameters optimization was conducted considering the loss definition given by Eq. (2). The data were split randomly into 3 sets: 80% of the configurations were used as training set, 10% were used as test set, which allows to evaluate how the model performs on unseen data during training, for a fixed set of hyperparameters, and 10% were used as validation set, to compare the accuracy of models trained with different hyperparameters. Both the Adam optimization algorithm and the mini-batch Stochastic Gradient Descent (SGD) one were tested, leading to similar results. The weights of the Neural Network were initialized randomly.

The following ranges were investigated for each parameter:
- Learning rate = $10^{-1} - 10^{-3}$
- Decay ratio = $10^{-1} - 10^{-2}$
- Batch size = $10^1 - 10^2$
- $\rho = 10^{-2} - 10^{-3}$ or $a = 0.1 - 0.8$
- $N_{cg}$ = from 2 up to half the number of atoms in the molecule

The optimal values chosen were those that yielded the lowest loss calculated on the validation set.

From the systematic testing of the various hyperparameter combinations, we observed significant variability in the obtained results, in terms of mappings produced, final values of the loss, and reconstruction accuracy. No systematic trends or correlations with the hyperparameter values seemed to emerge for the case of the learning rate, the batch size and the decay ratio. Some weak tendencies were observed: on average, a larger batch size and larger decay ratio values yielded lower loss values, therefore $10^2$ and $10^{-1}$ were adopted for the batch size and the decay ratio, respectively. Concerning the learning rate, provided that enough epochs were used, to allow the training to converge, we did not observe a clear effect on the final results, therefore a higher learning rate was used, for computational efficiency. The effect of the force regularization weight $\rho$ is discussed in detail in Section 3.2, as it has profound influence on the produced mappings. Apart from the hyperparameters values, it was noted that a source of significant variability of the results lies in the random initialization of the network weights, which will be further discussed in Section 3.3.

## 3.2 Effect of force regularization and loss normalization

The values of $\rho$ and $N_{cg}$ are tightly interconnected, as noted also by Wang et al. [6]. In particular, for the systems analyzed in this work, the reconstruction component of the loss typically reaches values of the order of $10^{-1}$ or $10^{-2}$, while the unscaled force component changes by orders of magnitude as the complexity of the molecule increases: ethane $\sim 10^1$, ethylbenzene $\sim 10^2$, CTA $\sim 10^3$, PIM-1 $\sim 10^3$ in units of $kcal^2/Å^2/mol^2$. Even though this difference across systems is expected, it adds difficulties in the selection of the parameter values and in the interpretation of the results.

Moreover, it is not clear what criterion should be used to guide the selection of an appropriate value for $\rho$, as lower values will always result in lower losses. It seems, though, important that the force regularization does not significantly exceed the reconstruction loss, otherwise this will trigger problematic behavior during the model training.

Especially in the cases where the force regularization is predominant in the loss function, the model progressively lumps together more atoms, effectively using fewer CG beads to perform the assignment than the total number available. This continues until the extreme case is reached, in which the majority of the encoder weights are zeroed and the whole molecule is mapped into a single CG bead. This behavior is independent of the VAE latent dimension size. Although this clearly minimizes the force component of the loss, it significantly worsens the reconstruction accuracy. If the model is intended to be used to perform backmapping, through the Decoder, this decrease in the reconstruction accuracy could be unacceptable.

An example of this behavior is shown in Fig. 2, for the case of PIM-1, using $\rho = 10^{-2}$. Even though the latent dimension of the VAE is 10, the model progressively reduced the number of beads effectively used in the assignment until only one was left. Around epoch 300, when the reduction from 2 to 1 used CG beads takes place, it can be observed that the reconstruction component of the loss indeed increases almost by a factor 3, compared to the lowest values reached during previous epochs. It is important to notice that, even if the model does not reduce to a single bead, depending on the value of $\rho$ it is not always possible to achieve the CG level desired, equal to the size of the latent dimension, since part of the latent variables will be discarded by the model. Moreover, it should be noted that, when the force regularization is present, the loss depends mostly on the number of used CG beads, rather than on how the atoms are subdivided among the moieties.

Since the value of $\rho$, which brings the force and reconstruction terms of the loss to a comparable level, is system-specific, the loss defined by Eq. (3) was introduced. With this normalization, the degenerate cases were $N_{cg}= 1$ were avoided and the need to search for an optimal value of $\rho$ for each system is bypassed. Indeed, the equivalent of $\rho$ in the normalized loss, $a$, did not have a systematic effect on the results for values of 0.1, 0.5, 0.8. More extreme values were not considered because they would defeat the purpose of the normalization, which is to bring the two components of the loss to the same order of magnitude. Therefore, we observed that $a$ could also be eliminated (which is equivalent to considering it equal to 0.5) and the model implementation could be simplified by including one less hyperparameter.

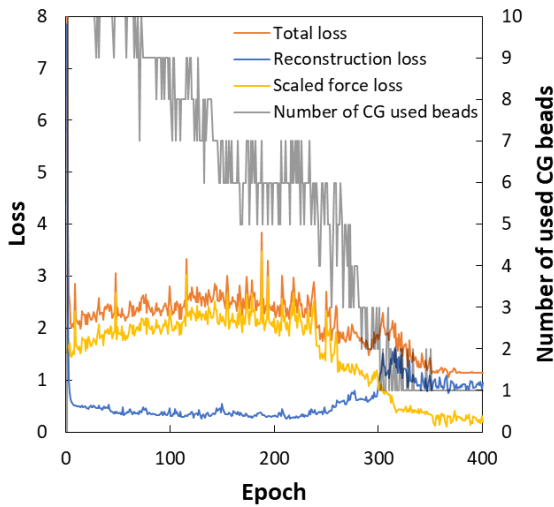

**Figure 2**: Trend of the total loss and of its reconstruction and scaled force components during training, for the case of PIM-1, using $\rho = 10^{-2}$ and $N_{cg} = 10$. On the right vertical axis, the effective number of used CG beads in each epoch is reported.

## 3.3 Multiplicity of mappings and connectivity preservation

An unexpected finding of this study is that the mapping obtained from the trained VAE model depends on the random initialization of the network weights. Training the model multiple times with the same hyperparameters yielded different, non-equivalent mappings. This was a general behavior, manifested for all hyperparameter combinations and for all the molecules tested, both simple and complex ones, as shown in **Fig. 3** for $C_2H_6$, and **Fig. 4** for PIM-1. In particular, in the latter case shown, not only the assignments, but also the number of CG moieties chosen by the model differ, which was a common occurrence.

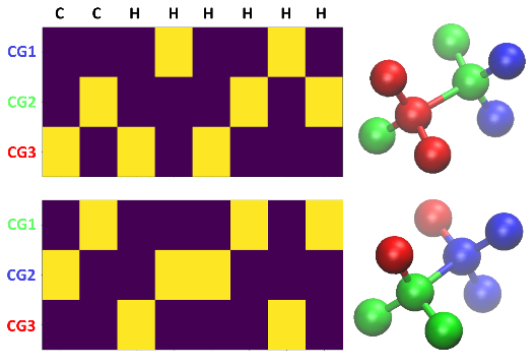

**Figure 3**: Assignment of $C_2H_6$ atoms to different CG moieties, indicated with yellow. Results obtained with the same hyperparameters and only varying the network initialization.

**Fig. 4** also displays another problematic feature encountered in several of the produced results: the VAE model sometimes groups together into the same CG bead atoms that are located far apart, thus breaking, in the CG representation, the link with the connectivity of the underlying atomistic model.

In order to avoid these unphysical mappings, at first an additional term was introduced into the loss function, which would penalize connectivity-breaking assignments. An assignment would be considered connectivity-preserving if, for all pairs of atoms belonging to a CG moiety, it would be possible to find a connecting path that did not involve atoms belonging to a different CG moiety. Although a promising concept, this loss function did not manage to reproduce interesting results. In particular, we observed that, in order to minimize the connectivity penalty, degenerate mappings of all atoms included into the same CG moiety were produced. Moreover, the evaluation of this loss function is significantly more costly, as it requires the creation of subgraphs for the atoms belonging to each moiety.

Therefore, this constraint was enforced instead by implementing an acceptability verification at the end of training, ensuring that only connectivity-preserving mappings were retained, and repeating the training with a different random initialization of the network until an acceptable solution was found. It should be noted that the presence or absence of the force regularization does not prevent the generation of connectivity-breaking mappings.

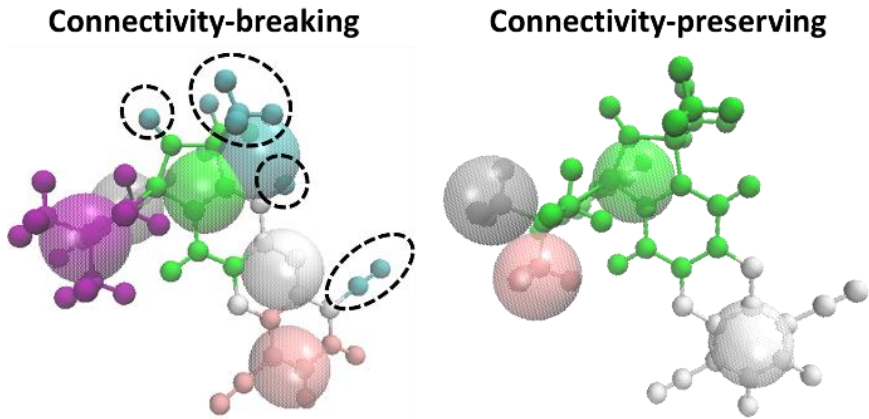

**Figure 4:** Example of connectivity-breaking and connectivity-preserving mappings produced for PIM-1. Results obtained with the same model hyperparameters and only different network initialization.

## 3.4 Reconstruction

### 3.4.1 Effect of diversity between configurations

As previously anticipated, when the force regularization is included in the loss, the reconstruction accuracy tends to decrease, therefore, in order to investigate aspects related to the possibility to use the Decoder to backmap from the CG space to the atomistic space, the following tests were conducted using the loss function defined by Eq. (1).

To test the approach on a $C_2H_6$ molecule, two models with latent dimension of 2 and 3 nodes were considered. Two CG beads correspond to the resolution obtained in the study by Wang et al. [6] for the same molecule. In **Fig. 5**, the trend of the loss function during the training can be observed, as well as the progression of the reconstructed molecule. For the 2-bead case, the model indeed converges to a chemically intuitive partitioning of the system, however, surprisingly, the reconstructed molecule lies on a single plane. Since a VAE model provides a lossy compression of the input, some information loss is to be expected. In the case of low dimensionality both in the input/output and latent space, the expressive power of the model might be insufficient to reconstruct the system in 3D.

On the other hand, with 3 nodes in the latent dimension, the VAE learns to accurately reconstruct the molecule. However, as will be discussed in detail also for ethylbenzene, this positive result is strongly dependent on the molecular configurations present in the dataset. In particular, accurate reconstruction hinges on the fact that the various configurations considered did not differ significantly from one another in their positions, and only a limited portion of the 3D space was explored within that dataset.

In the case of ethylbenzene, configurations with larger differences in terms of positions are present in the dataset. In **Fig. 6**, each colored region represents the positions in the 3D space sampled by one atom in the various molecular configurations present in the dataset. The upper part of the figure contains the whole dataset, while the lower part only the first $10^2$ configurations.

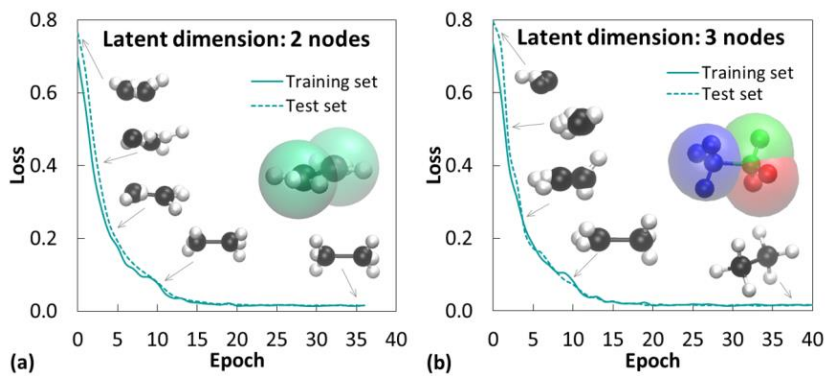

**Figure 5**: Loss over the training and test sets along with reconstructed molecules during training of the VAE for the ethane molecule with (a) 2 and (b) 3 latent nodes.

When the model is trained using all available configurations, the reconstruction of the aliphatic section of the molecule fails (Fig. 6a). In the dataset, those atoms explored a larger portion of the 3D space, rotating around the bond connecting to the aromatic ring. In the reconstruction, each atom is placed by the VAE in the average position that it had spanned in the dataset. As can be observed, this leads to a backmapped configuration with overlapping between atoms in an unrealistic fashion. Interestingly, the aromatic section is reconstructed well, and it can be seen how, for this section of the molecule, there was less significant displacement around the average position for each atom. If the model is trained with fewer, more similar configurations (Fig. 6b), a good reconstruction is obtained for the whole molecule. This behavior is exemplified here for ethylbenzene, but similar results were obtained also in the case of CTA and PIM-1.

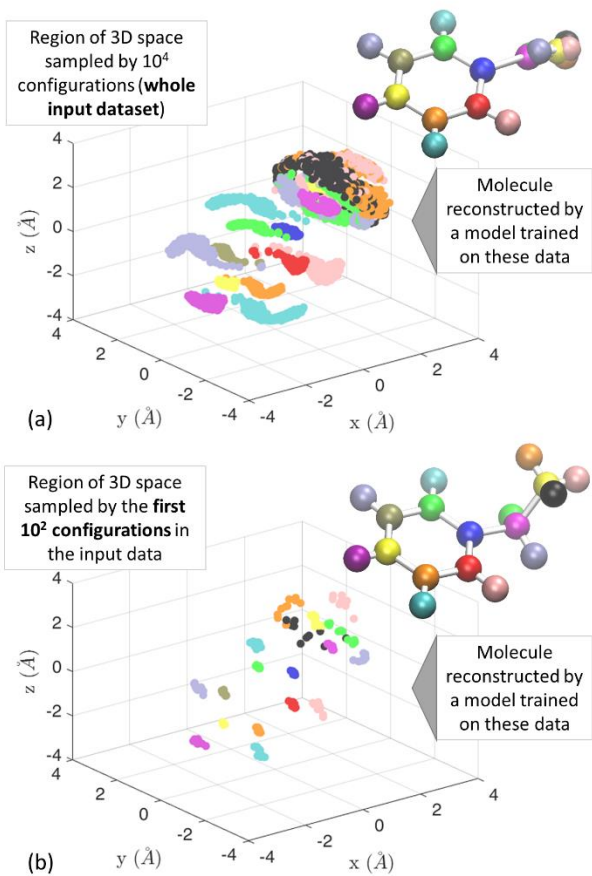

**Figure 6:** Relationship between the similarity among the positions in 3D space of the configurations used during training and reconstruction accuracy. VAE model trained with $N_{cg}$ = 2.

*3.4.2 Absence of rotational invariance*

A limitation encountered for the use of the VAE model – in the current implementation – to perform backmapping lies in the fact that it does not satisfy rotational invariance, as exemplified in **Fig. 7** for the case of CTA. In the upper part of the figure, it can be observed that a model trained with a small set of similar configurations (in the spirit of **Fig. 6b**) exhibits indeed high reconstruction accuracy. However, when the same model is applied to a rotated version of the same input configuration, it is unable to reconstruct it.

This behavior stems from the use of Cartesian coordinates as input features and suggests that the investigation of a rotationally invariant input feature, such as interatomic distances, to overcome this limitation should be pursued. In the current implementation, these limitations severely hinder the application of the VAE model to a multi-molecule system, or to a set of configurations (trajectory) where the molecules explore different positions, rotations, and conformations. Training on such data would produce a model that only reconstructs the average position of each atom, yielding unrealistic molecular conformations.

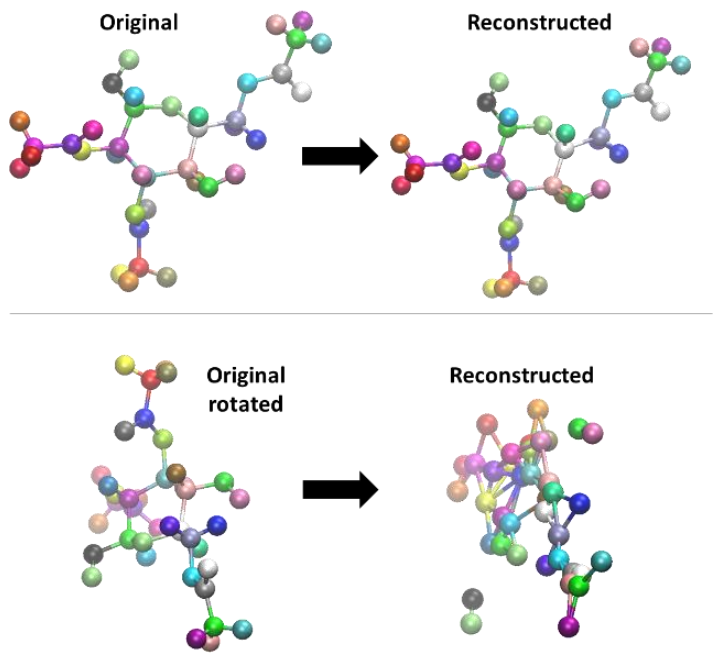

**Figure 7:** Effect of a rigid rotation of the input on the reconstructed output.
VAE model trained with $N_{cg} = 8$.

## 4 CONCLUSIONS

In this work, we investigated the use of a VAE to perform automatic mapping of atomistic systems into coarse grained beads, and reconstruct the atomistic detail, *i.e.*, backmapping. The model contains several hyperparameters, whose optimization is a costly step in terms of computational resources. In this architecture and for the systems studied, the learning rate, the batch size and the decay ratio did not appear to play a critical role on the results, therefore computational efficiency was the guiding criterion in the selection. Within the range of values tested, their effect seemed to be overshadowed by the effect of randomness in the network initialization. Therefore, with the exception of $\rho$ and $N_{cg}$, it would seem to be unnecessary to conduct an extensive search for optimal values of the VAE hyperparameters.

If a normalized definition of the loss function is introduced, it is possible to avoid the problematic behavior related to the strong effect of the force regularization, *i.e.* the collapse of all atoms into one single CG bead. Moreover, normalization avoids the necessity to introduce, and optimize, the force regularization weight as an additional hyperparameter. In fact, the same regularizing effect can be achieved by adopting different values of $N_{cg}$ (lower number of beads corresponds to smoother potential energy hypersurface in the CG space), combined with the filtering step of the output, that eliminates the connectivity-breaking mappings, which would introduce unfavorably high energy conformations of the CG molecule.

These observations possibly imply that the introduction of the force regularization is not necessary and that its avoidance has the added benefit of leading to higher accuracy of the reconstruction.

One of the more general observations of this study was the variety of results obtained, that did not exhibit any apparent correlation with the network hyperparameters and seemed to be a result of the different initializations of its weights. Evaluating which of the outputs would be preferable, requires considerations related to the intended use for the produced mapping, and might vary from system to system and from application to application. The VAE model, as modified in this study, can produce a variety of physically consistent options at various levels of resolutions, that can be further screened by devising fitness criteria for the specific application.

Concerning the reconstruction, in this implementation the model can only reproduce realistic configurations when trained on sufficiently similar input data, because, when a more diverse set is used in the training, only the average value is learned and reproduced in the reconstructed output.

From the analysis performed and the limitations highlighted, opportunities for future developments would concern the adoption of input features that preserve rotational and translational invariance during training. Moreover, as there is not a unique way to backmap the system from the CG to the atomistic level, the reverse mapping strategy should be extended via modifying the Decoder implementation in order to produce a variety of backmapped configurations, following the probability distribution of the underlying atomistic systems. The ultimate goal would be to associate the mapping procedure to the development of appropriate ML-based CG interaction potentials [7, 15, 16], that will enable the conduction of molecular simulations at the CG level and will provide a complete evaluation of the extracted mappings, in conjunction with the developed CG interaction potentials.

## ACKNOWLEDGMENTS

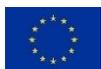
E.R. acknowledges funding from the European Union's Horizon 2020 research and innovation programme under the Marie Skłodowska-Curie grant agreement No 101030668. This work was supported by computational time granted from the National Infrastructures for Research and Technology S.A. (GRNET S.A.) in the National HPC facility - ARIS - under the projects COMPIL2 and MULTIPOLS (ID: 009014, 011032).